\documentclass[fleqn,10pt]{wlscirep}
\usepackage[utf8]{inputenc}
\usepackage[T1]{fontenc}
\title{Fast frequency discrimination and  phoneme recognition using a biomimetic membrane coupled to a neural network}

\author[1,2]{Woo Seok Lee}
\author[1]{Hyunjae Kim}
\author[3]{Andrew N. Cleland}
\author[1,*]{Kang-Hun Ahn}
\affil[1]{Department of Physics, Chungnam National University, Daejeon, 305-764, Republic of Korea}
 \affil[2]{Center for Theoretical Physics of Complex Systems, Institute for Basic Science (IBS), Daejeon 34051, Korea}
\affil[3]{Pritzker School of Molecular Engineering, University of Chicago, Chicago, Illinois 60637, USA}

\affil[*]{ahnkanghun@gmail.com}

\begin{abstract}
 In the human ear, the basilar membrane plays a central role in sound recognition. When excited by sound, this membrane responds with a frequency-dependent displacement pattern that is detected and identified by the auditory hair cells combined with the human neural system. Inspired by this structure, we designed and fabricated an artificial membrane that produces a spatial displacement pattern in response to an audible signal, which we used to train a convolutional neural network (CNN). When trained with single frequency tones, this system can unambiguously distinguish tones closely spaced in frequency. When instead trained to recognize spoken vowels, this system outperforms existing methods for phoneme recognition, including the discrete Fourier transform (DFT), zoom FFT and chirp z-transform, especially when tested in short time windows. This sound recognition scheme therefore promises significant benefits in fast and accurate sound identification compared to existing methods. 
\end{abstract}
\begin{document}

\flushbottom
\maketitle
%
%


\section*{Introduction}

The cochlea is the key organ for sound detection and recognition in mammals~\cite{Dallos,Robles}, comprising a mechano-electrical transducer that converts sound pressure into neuronal electrical signals. The cochlea generates these signals in a frequency-selective manner, due to the varying stiffness of the basilar membrane (BM) along the length of the cochlea~\cite{Ren}. For a given pure-tone audio signal, the location of the maximum displacement of the BM depends on the frequency of the tone: The base region responds to high-frequency sounds, while the apex region responds to low-frequency sounds. The audible range of the human auditory system, as determined by the mechanical response of the BM, ranges over twenty octaves, from roughly 20 Hz to 20 kHz, with a wide dynamic range of about 120 dB~\cite{Dallos}. 

Displacements of the BM are detected through the sympathetic motion of hair cells in the membrane, which generate electrical signals in the auditory nerves. The distinct spatial patterns of the BM's frequency-dependent response are encoded in the corresponding neural patterns, which are recognized in the brain. Using this auditory system, humans are able to accurately distinguish very short-duration sounds, while conventional frequency analysis faces significant challenges. 

Building a biomimetic analog of the BM, combined with an analog of the brain's pattern recognition system, provides an interesting approach for the frequency analysis of sound. Deep-learning techniques involving neural networks with multiple layers are important components in modern automatic speech recognition systems~\cite{Mohamed}, and have been used for feature representation~\cite{Hoshen}, acoustic modeling~\cite{Seide}, and language modeling~\cite{Arisoy}. In this study, we combine a physical model of the BM, yielding a frequency-dependent spatial response, with an artificial deep neural network for pattern recognition. 

The question we pose is what, if any, are the benefits of this approach to signal processing applications? Our findings can be summarized as follows:

i) A mechanical system with poor innate frequency selectivity can be significantly enhanced when combined with a trained neural network.

ii) The membrane together with the neural network provides enhanced pitch recognition for short duration signals, clearly distinguishing two frequencies $f_{1}$ and $f_{2}$ even when sampling for a time $T$ shorter than $1/|f_{1}-f_{2}|$. 

iii) The signal processing ability of the membrane combined with the neural network outperforms the standard discrete Fourier transform (DFT), the zoom fast Fourier transform (ZFFT) and the chirp Z-transform (CZT), the latter comprising popular methods for improving frequency resolution~\cite{lyons2004, porat1996, proakis2001}.
 
iv) We demonstrate that our system is helpful in recognizing real speech phonemes, and additionally can distinguish very short duration speech signals.

\section*{Membrane Fabrication and Experiments}
Our artificial basilar membrane (ABM) is a macroscopic structure mimicking von B\'{e}k\'{e}sy's model of the cochlea~\cite{Bekesy,Bekesy2}. Several research groups have developed similar approaches~\cite{White,Chen,Shintaku,Wittbrodt,Jung}, where frequency-selective ABMs have been designed in which the membrane width and thickness are varied along the membrane length, the response of these ABMs thereby mimicking the tonotopy of the cochlea. However, the motional amplitudes in these models are quite small, a few tens of nanometers for an 80--100 dB sound pressure level (SPL). Low-volume sounds are therefore quite challenging to detect using these structures, due to the difficulty of detecting very small (nm-scale) displacements. Here we demonstrate a new design for an ABM, which displays frequency-selective displacements for sound frequencies in the range of 100 to 1300 Hz, with displacements on the order of $10^{-4}$ m even for relatively small sound intensities of 65 to 70 dB SPL. This large responsivity makes detection of the sound-induced displacements relatively straightforward.

The ABM we used in this study, and the water-filled chamber it resides in, were designed in reference to von B\'{e}k\'{e}sy's model~\cite{Bekesy,Bekesy2}; a schematic is shown in the Supplementary Information (SI). Our ABM comprises a 0.1 mm thick sheet of silicone rubber, whose width varies along its 2 cm length from 1 mm to 5 mm. The membrane forms the mid-plane of a water-filled chamber, where the upper and lower parts of the chamber are analogs of the vestibular and tympanic canals, respectively. The upper and lower chambers are connected by a 5 mm diameter hole, analogous to the helicotrema in the cochlea. The upper part of the chamber is directly connected to an audio speaker with a plastic rod (4.5 cm), delivering an acoustic signal to the entire membrane. The sound pressure level was measured by driving the speaker with the same power and placing a microphone immediately below the ABM and recording the value. 

Acoustic signals emitted from the speaker produce spatial displacement patterns in the membrane, in synchrony with the audio signal, where the steady state response is reached in a time less than about 10 to 20 ms. The displacement pattern of the ABM is detected by reflecting beams of light from $n$ separate lasers (650 nm wavelength, Laserlab LDC650-3.5-5) onto $n$ one-dimensional position-sensitive detectors (PSDs, Hamamatsu Photonics S3932). We used $n=6$ for two-tone frequency discrimination and $n=10$ for vowel recognition.

\begin{figure} [hbt!]
\centering
\includegraphics[width=0.5\columnwidth]{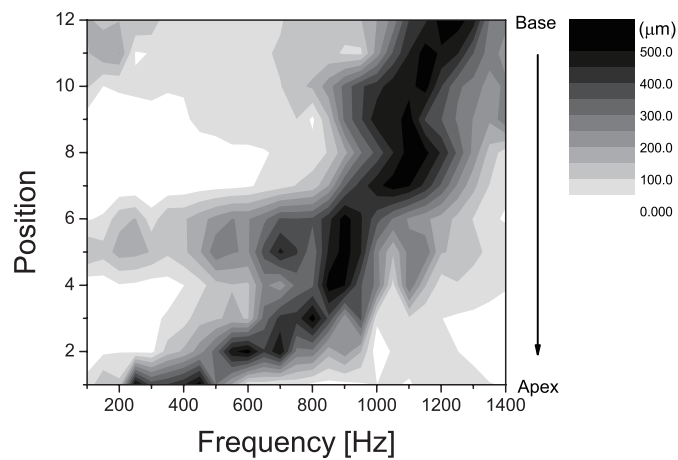}
\caption{\label{supple_fig2} Displacement response of the ABM (grey scale) as a function of position along the ABM (vertical scale) and the frequency of a continuous pure audio tone (horizontal scale), with displacement amplitudes measured after the ABM displacement has reached steady state. For this measurement we monitored the root-mean square (RMS) displacement amplitude by scanning 12 positions on the ABM using a single laser. Displacement position 1 is located 4 cm from the apex, and positions 2-12 are spaced evenly by 0.8 cm toward the base direction. 
\label{tonotopy}
}
\end{figure}

The pattern response of the ABM was measured by applying pure tone signals to the ABM with tone frequencies ranging from 100 Hz to 1300 Hz, at intervals of 100 Hz, with sound amplitudes of 65--70 dB SPL. The measured RMS displacement response is shown in Fig. \ref{tonotopy}. Low-frequency tones yield large displacements toward the apex end of the membrane, with the maximum deflection shifting towards the base with increasing frequency, a result of the variation in membrane width from apex to base; this is similar to the cochlear response. Unlike the cochlea, we have motion along the entire length for each frequency tone, possibly because our membrane lacks the feedback mechanism provided by the outer hair cells in the cochlea. Nonetheless, we can use the measured frequency-dependent spatial response to identify frequency components in a more complex audio signal.

\section*{Neural networks}

\begin{figure}[hbt!]
\centering
\includegraphics[width=\columnwidth]{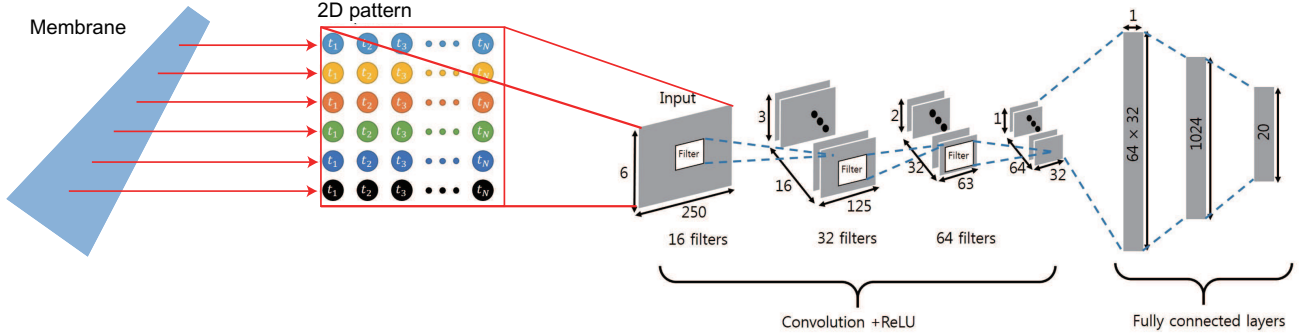}
\caption{\label{fig_system} Overview of the membrane-neural network system. Sound is applied to the artificial membrane (left) and the resulting movements are detected and converted into a 2D pattern (center), with rows representing membrane position and columns representing time series data. This pattern is used to train a CNN (right). The network structure depicted is 2D Network 1 in SI. 
}
\end{figure}

The neural network model we use to analyze the spatial patterns of the ABM is the convolutional neural network (CNN) \cite{LeCun} with a fully connected (FC) network \cite{krizhevsky}. The required nonlinear response of the system is provided by the rectified linear unit (ReLU) \cite{Nair}. For supervised learning, we use a quadratic cost function. An overview of the membrane-neural network system is shown in Fig.~\ref{fig_system}, and details of the implementation and results for the neural network training and testing are given in the SI.  

{\bf  Fourier transformation and other numerical methods for the enhancement of frequency resolution}

The standard numerical method for extracting frequency components $\tilde{s}(f_j)$ from a one-dimensional discrete times-series signal $s(t_k)$ is to use the discrete Fourier transform (DFT),
\begin{equation}\label{eq.four}
    \tilde{s}(f_j) = \frac{1}{N}\sum_{k=0}^{N-1} s(t_k)\exp(i 2\pi j k/N), 
\end{equation}
where $f_j = j \Delta f~(j=0 \ldots N-1)$ and $t_k = k \Delta t ~(k=0 \ldots N-1)$. The frequency resolution is limited to $\Delta f = 1/(N \Delta t) = 1/T$, where $T = N \Delta t$ is the total duration of the time series. 

Two methods that have been developed to improve the frequency resolution are known as the zoom fast Fourier transform (ZFFT) and chirp Z-transform (CZT). The principal concept for the ZFFT is to use the same total sample duration $T$, but to increase the time step $\Delta t$, resulting in down-sampling of the data. Any resulting non-integer relation between $T$ and $\Delta t$ is compensated by zero-padding. This gives an increased frequency resolution $\Delta f$ from the same data set, however accompanied by the loss of high-frequency components due to down-sampling, and the addition of spurious frequency components due to zero-padding. Increased down-sampling loses information about the original signal and reduces the quality of the transform.

The chirp Z-transform CZT is a generalization of the DFT. The DFT samples the transform plane at uniformly-spaced points along the unit circle, while the chirp Z-transform samples along spiral arcs, effectively using a non-uniform frequency coverage. It is defined as
\begin{equation}\label{eq.czt}
    \tilde{s}(f_j) = \sum_{k=0}^{N-1} s(t_k) A^{-j}W^{jk} 
\end{equation}
where $A=e^{i f_i/f_s}$ and $W=e^{-i 2\pi (f_f-f_i)/M}$, $f_i$ and $f_f$ are the initial and final and frequencies, and $M$ the number of samples in frequency. The chirp Z-transform can significantly enhance the frequency resolution without increasing the number of sampling data points, but it generates significant spectral leakage. For wider frequency signals, the result of the CZT is inaccurate.

Unlike these and other frequency refinement methods where no additional information is added to the time-series signal, the use of the mechanical response of the ABM in our approach adds spatial information to the time-series signal through the tonotopy, resulting in better frequency resolution without loss of signal. 

\section*{Training neural networks}
We trained the neural network using membrane patterns generated by combinations of two pure tones, using displacement data corresponding to time segments ranging from 20 ms to 200 ms in duration, with amplitudes of 65--70 dB SPL. The membrane displacement shows frequency selectivity for signals from 100 to 1300 Hz (SI). We used two pure tones, each with frequency in the range 600 to 695 Hz, with a step size and minimum frequency difference of 5 Hz, giving 190 different frequency combinations (cases). The network was trained so that each output node gives a binary output for each frequency component; training using two-tone signals, where two distinct nodes should yield TRUE outputs, therefore can discriminate 190 distinct frequency combinations, matching the number of cases. We trained the network with 200 data sets per case, and tested with 20 data sets per case, with no overlap between the training and test data. After training, the neural network yielded  high test accuracy (over 95\%) for the time widow size of 30--50 ms.

The training  errors are used to optimize the network parameters, using a back-propagation algorithm \cite{Rumelhart2}. A test error can occur when the trained network is operated with data outside the training set. We compared the training and test error rates when training the network with data from just one displacement channel to that using six displacement channels. With just one channel, the learning speed of the network was very slow, and with poor final  outcomes, with a highly-trained error rate of about 20\%. Using six displacement channels, however, yielded much faster training as well as significantly fewer training errors, as shown in the SI.

\begin{figure}[h]
\vskip 0.2in
\centerline{\includegraphics[width=0.8\columnwidth]{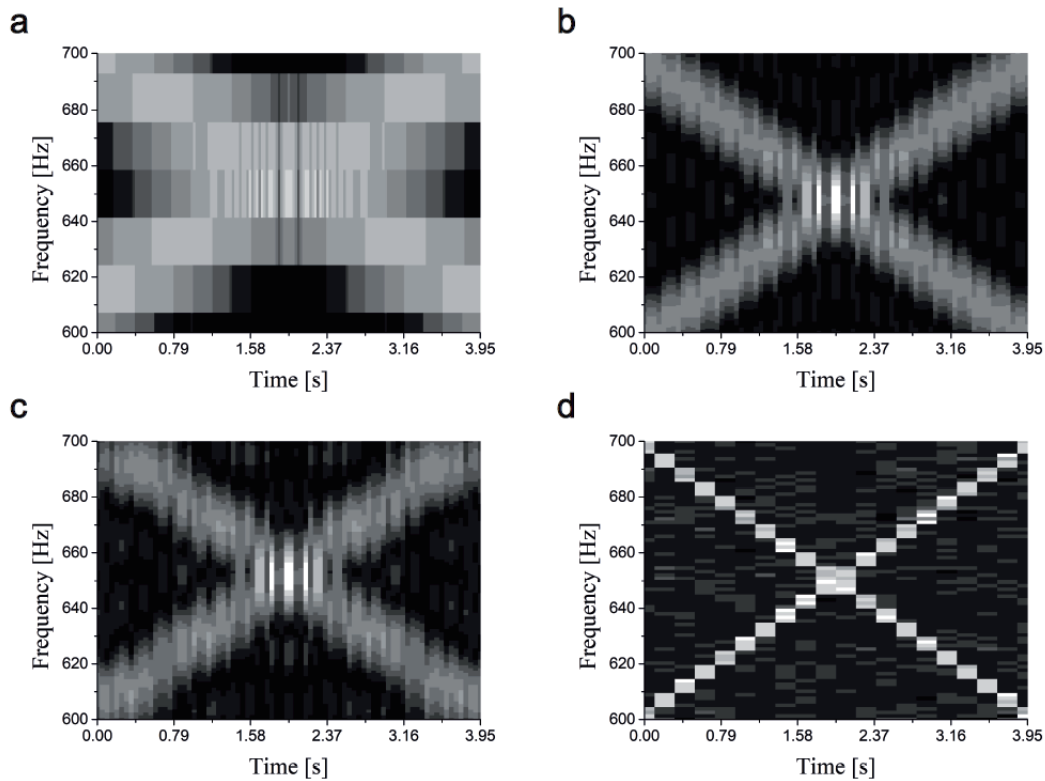}}
\caption{Analysis of signals with two frequency components of 50 ms duration, in 5 Hz intervals, (a) using DFT,
(b) using ZFFT, where
the input data is down-sampled with a sampling rate of $\frac{1}{20 \times \Delta t}$, and
(c) using CZT.
(d) Results using the ABM combined with the neural network analysis.}
\label{fig_sweep}
\vskip -0.2in
\end{figure}

\section*{Results}
We find that multi-channel signal processing using the artificial basilar membrane and the neural network provides a frequency resolution that is superior to existing methods, especially when using small sampling time windows.  In Fig.~\ref{fig_sweep}, we compare the frequency resolution for 50 ms-duration, two-tone signals, achieved using three numerical methods (DFT, ZFFT, and CZT) with that achieved with our trained network (2D network5 in the SI). The test data used to generate the neural network results in Fig.~\ref{fig_sweep}(d) are distinct from the data used to train the network. Our biomimetic system clearly resolves the two frequency tones, even for frequency differences as small as 5 Hz. By comparison, the resolution $\Delta f$ of the DFT is $\Delta f = 1/T=1/50~{\rm ms}=20~{\rm Hz}$ for a 50 ms time window, and the resulting poor resolution is shown in panel (a). Frequency components such as 605~Hz and 610~Hz cannot be resolved. The frequency resolution is improved with the CZT and ZFFT methods, as shown in Fig.~\ref{fig_sweep} (b) and (c), with the frequency resolution $\Delta f$ reduced to 5~Hz. However, in this case, about 30\% of CZT and ZFFT results have the wrong frequency peak values, with an error of about 5~Hz.

In contrast, when performing frequency analysis with the membrane-based pattern recognition, it is possible to resolve frequencies separated by only 5~Hz, smaller by a factor of four from the DFT result, with a frequency accuracy close to 100 percent.

\begin{figure}[h]
\vskip 0.2in
\centerline{\includegraphics[width=0.5\columnwidth]{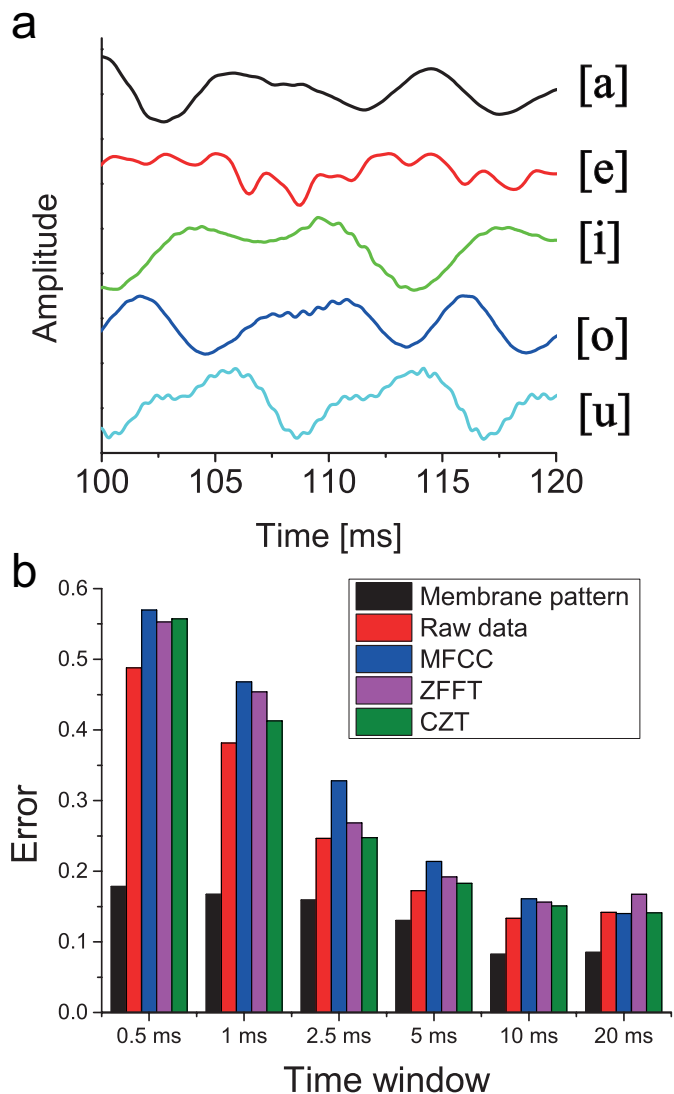}}
\caption{(a) Time series raw acoustic signals collected by a microphone, for the spoken vowels [a], [e], [i], [o] and [u]. (b) Test error rates of the trained CNN, using different time windows for the five vowels, using each of the five data pre-processing methods. The  pattern recognition error from the multi-position detection (black) are compared with the errors from raw time series data (red), conventional MFCC analysis (blue), ZFFT analysis (magenta), and CZT analysis (green). Horizontal axis shows the size of the time window.    }
\label{fig_vowel}
\vskip -0.2in
\end{figure}

In addition to the good frequency identification delivered by the ABM, we find that fast recognition of complex sounds is feasible using the same system, a problem that is significantly more complex than resolving a two-tone signal. We test the performance of the ABM combined with a trained neural network, with the CNN trained to recognize five different Korean vowels with different sampling time windows (0.5, 1, 5, 10, 20 ms). We use the voice data sets from Ref.~\cite{kim2019origin}, with 60 voices as training data and 15 voices as test data (Fig.~\ref{fig_vowel} (a)).

We compare the vowel recognition rate of a convolution neural network using different inputs to train the CNN. These include the ABM pattern as well as a number of other common methods used to identify spoken sounds.  These include using the raw data (with no pre-processing) as the CNN input, the outputs of the CZT and ZFFT transforms, and a conventional Mel-frequency cepstral coefficient (MFCC) analysis. The MFCC method provides a representation of the power spectrum of a sound, using a cosine transform of the log power spectrum on a nonlinear Mel frequency scale \cite{sahidullah}. For the MFCC and raw data recognition, we used the neural network structure in  Ref.~\cite{kim2019origin}; using this, we achieved a 91\% accuracy for a 50 ms time window. For the CZT and ZFFT analyses, we set the frequency band to 200 Hz to 3000 Hz in order to cover most of the vowel frequency components. When using the ABM to transform sound, we measured the ABM membrane displacement patterns using ten lasers, where position 1 is located 4 cm from the apex, and positions 2--10 are spaced evenly by 0.8 cm toward the base. The CNN is trained for each time window with a unique data set, then tested with different data.

As shown in Fig.~\ref{fig_vowel} (b), we find that all methods give roughly equivalent performance when using relatively long time windows (5--20 ms), while the performance of the ABM displacement window is clearly superior for shorter time windows, especially below 2.5 ms. 

We find that as the vowel sampling time window decreases, the accuracy of each of the standard methods decreases drastically, to about 50\%, while the ABM pattern accuracy is only slight reduced, from 90\% to 80\%. 
 
When the Fourier components of a speech signal are analyzed in a short time window, the frequency resolution is poor due to the resultingly poor frequency resolution. The ABM pattern recognition sidesteps this limitation, because the ABM pattern recognition involves both spatial and temporal information, where the spatial response is convolved with some frequency information, even for very brief excitations.
 
To understand qualitatively why the mechanical response pattern improves the performance, we consider a simplified model of the membrane as a harmonic oscillator with mass $m$, friction constant $\gamma$, and angular resonance frequency $\omega_{0}$. The displacement of the oscillator $x(t)$ in response to a driving force $F(t)$ follows the Newtonian equation of motion, with displacement
\begin{eqnarray}
 x(t)=\int _{-\infty}^{t} dt^{\prime}G(t-t^{\prime}) F(t^{\prime})/m,
\end{eqnarray}
where the Green function response $G(t-t^{\prime})$ is given by
\begin{eqnarray}
 G(t-t^{\prime})=\frac{1}{\pi}\exp(-\frac{\gamma}{2}(t-t^{\prime}))\frac{\cos \left [\sqrt{4\omega_{0}^2-\gamma^{2}}(t-t^{\prime})\right ]}{\sqrt{4\omega_{0}^{2}-\gamma^{2}}   }.
\end{eqnarray}
Note that $G \propto \exp(-\gamma(t-t^{\prime})/2)$, so the displacement $x(t)$ at time $t$ integrates the sound over a preceding time window $\sim 2/\gamma$. In the ABM, measurements of the mechanical $Q$ factor indicate that $2/\gamma\sim$ 1~ms, so the displacement $x(t)$ of the oscillator encodes roughly that length of the sound signal.

The integrated response is however not sufficient to explain the dramatic reduction of the error rate of the ABM for short time windows, shown in Fig.~\ref{fig_vowel}. The ABM membrane's motion is however sensed in multiple locations, each of which can be roughly modeled as an independent harmonic oscillator, with a different resonance frequency. The sound identification involves the pattern from many (in our experiment, $n=10$) such oscillators. In auditory science, this position-dependent frequency analysis is called ``place coding.'' The ABM-enabled place coding used here, combined with the known strong pattern recognition of our neural network, is what enables the good sound recognition capability demonstrated in this work.

\subsection*{Conclusion}
We have demonstrated an artificial basilar membrane, with a frequency-dependent spatial response, which when combined with a trained pattern-recognizing neural network shows outstanding performance in frequency resolution as well as speech recognition of short-time phonemes. This artificial system  mimics the human auditory system's ability to resolve and distinguish very short segments of speech, and shows strong potential for analysis of more complex spoken sounds. 


\section*{Acknowledgements}
We thank Jong Hun Park for the information on CZT. This work was supported by the National Research Foundation of
Korea (NRF) grant funded by the Korea government (MSIP), NRF2017R1A2B3010002.

\section*{Author contributions statement}
K.H.A. and A.C. designed the research, W.S.L. and H.K. performed experiments. K.H.A. and H.K. analyzed data. W.S.L., A.C., and K.H.A. wrote the paper.

\section*{Additional Information}
\textbf{Competing Interests}: The authors declare no competing interests.

\end{document}